\documentclass[prd,a4paper,twocolumn,preprintnumbers,nofootinbib,superscriptaddress]{revtex4}
\usepackage[english]{babel}
\usepackage{amsmath,amssymb,amsfonts, bm,bbm,slashed, subdepth}
\usepackage{graphicx}
\usepackage{enumerate}
\usepackage{setspace}
\usepackage{booktabs, tabularx}
\usepackage{units}
\usepackage{color}
\usepackage{multirow}
\usepackage[dvipsnames]{xcolor}
\usepackage[pscoord]{eso-pic}
\usepackage[normalem]{ulem}

\usepackage{scalerel}
\usepackage{soul}

\usepackage{hyperref}
\hypersetup{ 
	setpagesize=false,
	bookmarksnumbered=true,
	colorlinks=true,
	linkcolor=blue,
	citecolor=red,
	hypertexnames=true
}

\newcommand{\be}{\begin{equation}}
	\newcommand{\ee}{\end{equation}}
\newcommand{\ba}{\begin{array}}
	\newcommand{\ea}{\end{array}}
\newcommand{\bea}{\begin{eqnarray}}
	\newcommand{\eea}{\end{eqnarray}}
\newcommand{\balg}{\begin{align}}
	\newcommand{\ealg}{\end{align}}
\newcommand{\bit}{\begin{itemize}}
	\newcommand{\eit}{\end{itemize}}

\definecolor{bostonuniversityred}{rgb}{0.8, 0.0, 0.0}

\makeindex

\begin{document}
	\preprint{ULB-TH/21-04}
	
	\title{
		Seesaw determination of the dark matter relic density
	}
	
	\author{Rupert Coy}
	\affiliation{Service de Physique Th\'eorique, Universit\'e Libre de Bruxelles, Boulevard du Triomphe, CP225, 1050 Brussels, Belgium}
	\author{Aritra Gupta}
	\affiliation{Service de Physique Th\'eorique, Universit\'e Libre de Bruxelles, Boulevard du Triomphe, CP225, 1050 Brussels, Belgium}
	\author{Thomas Hambye}
	\affiliation{Service de Physique Th\'eorique, Universit\'e Libre de Bruxelles, Boulevard du Triomphe, CP225, 1050 Brussels, Belgium}

\begin{abstract}
	In this article we show that in the usual type-I seesaw framework, augmented solely by a neutrino portal interaction, the dark matter (DM) relic density can be created through freeze-in in a manner fully determined by the seesaw interactions  and the DM particle mass. This simple freeze-in scenario, where dark matter is not a seesaw state, proceeds through slow, seesaw-induced decays of Higgs, $W$ and $Z$ bosons. We identify two scenarios, one of which predicts the existence of an observable neutrino line. 
	\end{abstract}
	
	\maketitle
	
The nature of DM as a particle and the origin of neutrino masses constitute two of the main conundrums of particle physics today.
Whether these two enigmas could be closely related is a fascinating question.
In the type-I seesaw scenario, the right-handed Majorana neutrinos and their Yukawa interactions 
allow for a particularly simple and motivated explanation of neutrino masses \cite{Minkowski:1977sc,Yanagida:1979as,GellMann:1980vs,Mohapatra:1979ia}.
In this article, we are interested in the possibility that these interactions could play an important role in the existence of DM today.
If the right-handed neutrinos lie around the electroweak scale or below, these seesaw interactions are expected to be small, so that they induce sufficiently suppressed neutrino masses. Thus, in this case, if these interactions are to play an important role in the production of DM, one would expect that it is  through out-of-equilibrium freeze-in production \cite{McDonald:2001vt,Hall:2009bx} rather than freeze-out production (i.e.~exit from thermal equilibrium). This possibility, where DM is slowly produced out of equilibrium in the early Universe thermal bath through processes where the small seesaw Yukawa couplings are involved, has been considered in several recent works \cite{Aoki:2015nza,Becker:2018rve,Bandyopadhyay:2020qpn,Ma:2021bzl,Chianese:2021toe}.
Putting aside the possibility of sterile neutrino DM (see e.g.~\cite{Dodelson:1993je,Shi:1998km,Boyarsky:2018tvu,Lucente:2021har} for studies of the constraints on this scenario), this requires an interaction between the right-handed neutrino(s) and DM. The most minimal possibility is to assume a neutrino portal Yukawa interaction
where the right-handed neutrinos couple to new scalar and fermion particles, one or both constituting the DM. 
In this article, we point out that based on this simple seesaw/neutrino portal structure, the DM relic density could be produced from Higgs, $W$ and $Z$ boson decays via freeze-in in a manner that depends only on the seesaw parameters and the mass of the DM particle(s).

\section{General setup}
	
We begin by displaying the Lagrangian assumed. On top of the usual type-I seesaw interactions, 
\begin{eqnarray}
	\hspace{-0.2cm}
	{\cal L}_\text{seesaw}&=&i \overline{N_R}\partial \hspace{-1.5mm} \slash N_R-\frac{1}{2} m_N (\overline{N_R} N_R^c+\overline{N_R^c} N_R)\nonumber\\
	&& -(Y_\nu \overline{N_R} \tilde{H}^\dagger L+h.c.)\, ,
	\label{seesaw}
	\end{eqnarray}
one assumes only a neutrino portal interaction, 
\begin{equation}
	\delta{\cal L} =  -Y_\chi \overline{N} \phi \chi  +h.c.\,.
	\label{LagrchiN}
\end{equation}
A sum over the various right-handed neutrinos is implicit.
We assume that $\chi$ is a two-component Majorana spinor. The generalisation to a four-component Dirac spinor is straightforward. We do not assume any symmetry at this stage, i.e.~$\chi$ and $\phi$ are singlets of all existing symmetries. 
The possibility that they are charged under a discrete, global or local symmetry will be discussed later.
	
As is well known, in the seesaw mechanism the neutrino masses follow from the diagonalisation of the induced neutrino mass matrix, ${\cal M}_\nu= - Y_\nu^T m_N^{-1}Y_\nu v^2/2$, where $v=246$ GeV is the vev of the SM scalar boson.
Given the value of the atmospheric and solar neutrino mass splittings, this implies that two right-handed neutrinos, which we will call $N_{2,3}$, necessarily have Yukawa couplings much larger than the typical $10^{-10}$-$10^{-13}$ values one needs to produce the observed relic density through freeze-in. 
However, one of the three right-handed neutrinos, which we will call $N_1$, or simply ``$N$'', could nevertheless have smaller couplings as the absolute neutrino mass scale is not known, i.e. the value of the lightest neutrino mass could be very tiny or even vanishing. 
If this right-handed neutrino is lighter than the Higgs boson, the Higgs boson can decay to $N+\nu_i$ (i$=$e,$\mu$,$\tau$) with a decay width\footnote{We will assume the two other right-handed neutrinos are heavier, $m_{N_{2,3}}> m_h$, and have negligible neutrino portal interactions. These could, for instance, be responsible for successful baryogenesis through leptogenesis (and without much washout of the $L$ asymmetry produced by $N_1$ interactions, given the smallness of the $N_1$ interactions and possible flavour effects). Alternatively, they could also never have been produced if the inflation reheating temperature is smaller than their masses.}
\begin{equation}
	\Gamma_{h\rightarrow \overline{N}\,\nu_i+N\bar{\nu}_i}=\frac{1}{16\pi} m_h |Y_{\nu i}|^2 \Big(  1-\frac{m_{N}^2}{m_h^2}\Big)^2 \,. 
\end{equation}
Similarly, in the electroweak broken phase, the decays of the $W^\pm$ to $N+l^\pm$ and $Z$ to $N \nu$ occur through $N$-$\nu$ mixing, if kinematically allowed, 
\begin{eqnarray}
	\Gamma_{W^\pm \rightarrow N\,l_i^\pm} &=& \frac{1}{48\pi} m_W |Y_{\nu i}|^2\, f(m_{N}^2/m_W^2) \, ,  \\
	\Gamma_{Z \rightarrow \overline{N}\,\nu_i+N \bar{\nu}_i}&=& \frac{1}{48\pi} m_Z |Y_{\nu i}|^2\, f(m_{N}^2/m_Z^2) \,, 
\end{eqnarray}
where $f(x) = (1-x)^2(1+2/x)$. 
Note the $m^2_{W,Z}/m_N^2$ enhancement (due to $N$-$\nu_i$ mixing) of the gauge boson decay widths with respect to the $h$ one.
Given the neutrino mass constraints, it is perfectly possible that these decays have never been in thermal equilibrium. 
For the $Z$ decay, which is the fastest, this requires 
$\Gamma_{Z\rightarrow \overline{N}\,\nu+N\bar{\nu}_i}/H|_{T\simeq m_Z}\lesssim 1$, i.e.
\begin{equation}
	\sum_i|Y_{\nu i}|^2 \lesssim 1 \cdot 10^{-16}\cdot \Big(\frac{m_N}{10\,\hbox{GeV}}\Big)^2\,,
	\label{Ynunonthermal}
\end{equation}
or
\begin{equation}
	m_{\nu_1}\leq \tilde{m}_1< 3\cdot 10^{-4}\,\hbox{eV} \, 
	(m_N/10\,\hbox{GeV}) \, ,
\end{equation}
where we have neglected $m^2_{N}/m_Z^2$ corrections and have used the well-known seesaw inequality 
$m_{\nu_1}  \leq  \tilde{m}_1\equiv \sum_i|Y_{\nu i}|^2v^2/(2 m_{N})$, with $m_{\nu_1}$  the mass of the lightest SM neutrino. 
As is also well known, the value of $\tilde{m}_1$ is experimentally allowed to be anywhere between $m_{\nu_1}$ and values much larger than the neutrino masses but, barring cancellations between the Yukawa couplings in the neutrino mass formula, it is expected below the upper bound on neutrino masses $\sim 1.1$~eV \cite{Schluter:2020gdr}, typically of order $m_{\nu_1}$. 
	
If $N$ has never been in thermal equilibrium (and not created at the end of cosmic inflation), then it can only be created through freeze-in. For $m_{N}\lesssim m_{h,W,Z}$, the dominant freeze-in production mechanism is from the above decay channel(s).\footnote{Scattering processes involving two powers of $Y_\nu$ in the amplitude clearly have a very suppressed contribution. Scattering processes from SM fermions involving only one power, e.g. $f\bar{f} \rightarrow N L$,  bring a smaller contribution ($\sim 20\%$) than the decays and for clarity we will limit ourselves to the contribution of the decays. Three-body SM fermion decays, $f\rightarrow {f}' \bar{L} N$, also give a subleading contribution.}
The resulting number of right-handed neutrinos from $Z$ boson decays is given by
\begin{equation}
	Y_{N}\equiv\frac{n_{N}}{s}= c_Z \cdot \frac{\gamma_{Z\rightarrow \overline{N} \nu+N\bar{\nu}}}{sH} \Big|_{T=m_Z} \, .
	\label{N1yield}
\end{equation}
Here we have simply multiplied the 
number of decays per unit time and volume, $\gamma_{Z\rightarrow \overline{N} \nu+N\bar{\nu}}$, by the age of the universe $\sim 1/H$, everything taken at about the freeze-in production peak temperature, $T\sim m_Z$, with
\begin{align}
	\gamma_{Z\rightarrow \overline{N} \nu+N\bar{\nu}}&=\langle n_{Z}^{eq}\Gamma_{Z\rightarrow \overline{N} \nu+N\bar{\nu}}\frac{E}{m_{Z}}\rangle \notag \\
	&=  \frac{m_Z^3 T}{32\pi^3} K_1(m_Z/T) f(m_N^2/m_Z^2) \sum_i |Y_{\nu i}|^2 \, , 
\end{align}
which depends on the Bessel function $K_1$, and where the brackets refer to the thermal average. 
This is valid up to a constant, $c_Z$, of order unity, cf. Eq.~\eqref{N1yield}. 
Integrating the corresponding Boltzmann equation gives $c_Z=3\pi/[2K_1(1)] = 7.8$.
The same formula holds for $W\rightarrow N l_i$ and $h\rightarrow N \nu_i$, and we find $c_W=c_h=c_Z$.
	
Once produced this way, the right-handed neutrinos can decay dominantly to $\chi+\phi$ through neutrino portal interactions, provided $m_{N} > m_\chi+m_\phi$ (see below for possible effects of three-body decays). This simply gives
\begin{equation}
	Y_\chi=Y_\phi=Y_{N} \, .
\end{equation}
Summing the contributions from $Z$, $W$ and $h$ decays, one finally obtains
\begin{equation}
	\Omega_{DM} h^2 \simeq
	10^{23} \, \sum_i |Y_{\nu i}|^2\, \Big(\frac{m_\chi+m_\phi}{1\,\hbox{GeV}}\Big)\,\Big(\frac{10\,\hbox{GeV}}{m_N} \Big)^2\,. 
	\label{Omegafreezein}
\end{equation}
Note that we have assumed here that both $\chi$ and $\phi$ are stable, so that both are DM components. If one of these particles is unstable, one has to drop the corresponding mass from this equation. Thus, one gets the observed value $\Omega_{DM}h^2=0.12$ if
\begin{equation}
	\sum_i |Y_{\nu i}|^2\simeq 10^{-24} \cdot \Big( \frac{m_N}{10\,\hbox{GeV}}\Big)^2\Big(\frac{1\,\hbox{GeV}}{m_\chi+m_\phi}\Big) \, , \label{Ynufreezein}
\end{equation}
which implies
\begin{equation} 
	m_{\nu_1}< \tilde{m}_1= 4 \cdot 10^{-12}\,\hbox{eV} \cdot \frac{10 \hbox{ GeV}}{m_{N}}\cdot \left( \frac{1\,\hbox{GeV}}{m_\chi+m_\phi} \right) \,. 
	\label{numass}
\end{equation}
Again, $m_{\nu_1}\simeq \tilde{m}_1$ holds approximately, unless there are cancellations between various Yukawa couplings in the neutrino mass formula.
Thus, the DM relic density is determined only by its mass and the seesaw parameters. In particular, one finds an interesting relation between the value of the lightest neutrino mass and the DM relic density. 
Such a tiny neutrino mass value would be very difficult to probe, but is falsifiable from absolute neutrino mass scale and neutrinoless double beta decay experiments such as KATRIN and GERDA \cite{Ackermann:2012xja}, as well as from cosmology.

\section{DM stability}
This very simple mechanism above has to be confronted with several constraints, which we will now discuss. 
Firstly, one has to check that the interactions assumed above 
do not lead to too fast a DM decay. There are different possibilities depending on the existence of various possible symmetries. Before discussing these, we will first limit ourselves to the decay which must occur simply as a result of the interaction assumed above, irrespective of the symmetry assumed.
	
\begin{figure}[t]
	\centering
	\includegraphics[width=0.85\columnwidth]{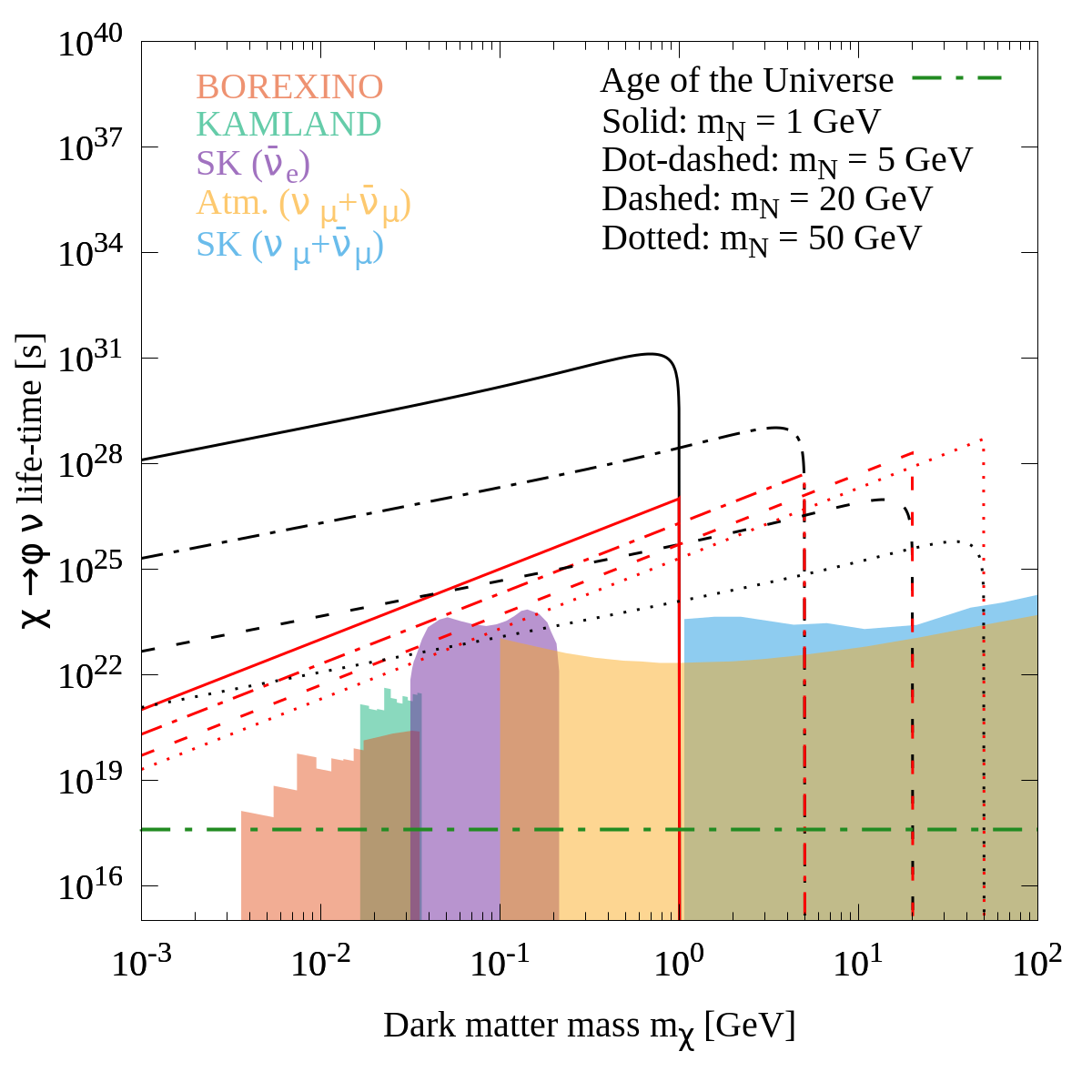}
	\vspace{-0.3cm}
	\caption{Constraints on neutrino line production~\cite{Collaboration:2011jza,Zhang:2013tua,Malek:2002ns,Bellini:2010gn,FrankiewiczonbehalfoftheSuper-KamiokandeCollaboration:2016pkv}. For various values of $m_N$, experimental lower bounds (see \cite{Garcia-Cely:2017oco,Coy:2020wxp}) are contrasted with the theoretical upper bound on the lifetime of $\chi\rightarrow \phi \nu$ decay, assuming that $m_\chi \gg m_\phi$, so that the neutrino line occurs at an energy equal to $m_\chi/2$. 
	The black lines denote the value of the lifetime below which the two-body decay channel is dominant, so that the seesaw/DM relic density correspondence holds. The red lines give the upper bound from structure formation constraints, Eq. \eqref{structureformation}.}
		\label{plot1}
\end{figure}

If $m_\chi>m_\phi$, the $\chi$ particle can decay in several ways. At energies of order or below the electroweak scale, as is the case here, the dominant decay is into $\nu+\phi$ via the $Y_\chi$ interaction and $\nu-N$ seesaw-induced mixing,\footnote{The $m_\phi<m_\chi$ case is very similar to the $m_\chi<m_\phi$ case, just inverting $\chi$ and $\phi$.  The relevant decay width is $\Gamma_{\phi\rightarrow \chi\nu}=\frac{1}{16 \pi} m_\phi \sum_i |Y_\chi|^2 |Y_{\nu i}|^2 v^2/m^2_{N}$,
leading to the same possibility of indirect detection
and about the same constraint on the $Y_\chi$ coupling.}
\begin{equation}
	\Gamma_{\chi\rightarrow \phi\nu}=\frac{1}{32\pi} |Y_\chi|^2 \frac{ \sum_i|Y_{\nu i}|^2 v^2}{m^2_{N}}m_{\chi}\Big(1-\frac{m_\phi^2}{m_\chi^2}\Big)^2 \, .
		\label{GammaChi}
\end{equation}
Here there are two options, depending on whether the 
$\chi$ lifetime is larger or smaller than the age of the universe.
We will mainly consider the former option (A), and also briefly discuss the latter option (B).
Option A has the nice feature of producing a monochromatic flux of neutrinos, i.e.~a DM smoking gun neutrino line. 
Note that since we consider $m_\chi$ up to the EW scale, this neutrino line can be much more energetic than, for instance, the case of keV sterile neutrino DM \cite{Dodelson:1993je,Shi:1998km,Boyarsky:2018tvu,Lucente:2021har}. 
The current lower bound on the lifetime of DM decays producing a neutrino line is given in Fig.~\ref{plot1}. For $m_{DM} \sim$ GeV, it is of order $10^{24}$~sec. Such a long lifetime requires, on top of the large $|Y_\nu|^2$ suppression of the decay width (see Eq.~\eqref{Ynufreezein}), an approximately equal suppression from $|Y_\chi|^2$,
\begin{equation}
	|Y_\chi|^2\lesssim 10^{-25}\,\Big(\frac{10^{24}\,\hbox{sec}}{\tau_{\chi}^{\rm{obsv}}}\Big) \, .
		\label{Ychiupperbound}
\end{equation} 
Approximately saturating this bound leads to an observable neutrino line.
This small value of $Y_\chi$ implies that the decay of $N$ is quite slow,
\begin{eqnarray}
	\label{GammaN2body}
	\Gamma_{N\rightarrow  \chi\phi}&\simeq &\frac{1}{16\pi}m_{N} |Y_\chi|^2\left(1+\frac{2\,m_\chi}{m_N}\right)\\
	&&\lesssim 3\cdot 10^{-2}\,\hbox{sec}^{-1} \cdot \Big(\frac{10^{24}\,\hbox{sec}}{\tau_{\chi}^{{\rm obsv}}}\Big)\Big(\frac{m_N}{10\,\hbox{GeV}}\Big)\nonumber \, ,
\end{eqnarray}
where in the inequality we neglected corrections of order $(m_{\chi,\phi}/m_N)$.
Note, importantly, that even if suppressed in this way, the $N\rightarrow \chi \phi$ decay width driven by $Y_\chi$ can still easily dominate over the various three-body decays that are induced by the $Y_\nu$ couplings, $N\rightarrow \nu f \bar{f}$ and $N \to \ell f \bar{f'}$. Thus, there is indeed a whole range of parameter space for which the one-to-one seesaw/DM relic density relation, Eq. \eqref{Omegafreezein}, holds. To see this, one has to compare the two-body decay width with the three-body one, which for the neutrino channel is
\begin{equation}
	\Gamma_{N\rightarrow\nu f \bar{f}}=\frac{N_c}{1536\,\pi^3}\,|Y_{\nu i}|^2\frac{g_2^2}{\cos \theta_W^2} (g_L^2+g_R^2)\frac{m_{N}^3}{m_Z^2} \, ,
\end{equation}
and similarly for $N \to \ell f \bar{f'}$.
As usual, $g_{L,R}=T_3-Q\,\sin^2 \theta_W$, with $g_2$ and $\theta_W$ the $SU(2)_L$ gauge coupling and Weinberg mixing angle, and $T_3$, $Q$, $N_c$  the weak isospin, electric charge, 
and number of colours of the SM fermion considered. 
Summing the three-body decays over all quarks and leptons allowed in the final state, 
the two-body decay width dominates when $|Y_\chi|^2\gtrsim 10^{-4} \,\sum_i|Y_{\nu i}|^2 \,(m_{N}/10\,\hbox{GeV})^2 $. This lower bound must be compared with the upper bound on $Y_\chi$ such that it doesn't induce too intense a neutrino line, Eq.~\eqref{Ychiupperbound}. 

The black lines in Fig.~\ref{plot1} display for various values of $m_N$ the value of $\Gamma_{\rm min,\chi\rightarrow \phi \nu}^{-1} \propto m_\chi/m_N^4$ below which the two-body decay dominates and hence for which there is the one-to-one correspondence.
To a large extent, the region where the correspondence holds 
predicts a neutrino line that we may hope to detect soon (except for $m_\chi$ well below GeV, where the experimental lower bound on the lifetime is less stringent).
Note, importantly, that if $m_{N}> m_{W,Z,h}$, one can show that unless $m_\chi$ is below $\sim \mathcal{O}(10)$\,MeV, the one-to-one correspondence between the seesaw parameters and the DM relic density is lost because in this case $N$ decays much faster through two-body decays into a SM lepton and a SM boson. In this case, freeze-in works through scattering processes \cite{Becker:2018rve,Bandyopadhyay:2020qpn,Chianese:2021toe,Cosme:2020mck,Chianese:2018dsz}.

The lower bound on the lifetime of $N$, Eq.~(\ref{GammaN2body}), may be a few orders of magnitude larger than the age of the universe at the BBN epoch, $\tau_{BBN}\sim 1-100$~sec. 
One could therefore wonder if BBN is a matter of concern. However, it is not the case because the number of $N$ particles decaying is very limited, and they negligibly contribute to the total energy density at this time (hence to the Hubble expansion rate),  even if $N$ decays into two particles which are relativistic.
Moreover, the decay is into $\chi$ and $\phi$, which do not cause any photo-disintegration of nuclei since they do not produce any electromagnetic or hadronic material, (unless the $\phi$ scalar has a vev and decays through a Higgs portal).
The late $N$ decay, producing relativistic DM, is nevertheless a matter of concern for structure formation. 
Imposing that DM, which has kinetic energy $\sim m_N/2$ when produced from $N$ decays, Eq.~\eqref{GammaN2body}, redshifts enough so that it is non-relativistic when $T\sim $~keV (so that it doesn't affect structure formation too much \cite{Irsic:2017ixq}) gives an upper bound on the $\chi$ lifetime,
\begin{equation}
	\tau_\chi \lesssim 10^{28}\,\hbox{sec}\,\Big(\frac{m_{DM}}{m_N}\Big)^2\Big(\frac{m_N}{10\,\hbox{GeV}}\Big) \, .
	\label{structureformation}
\end{equation}
We show the corresponding constraint in red in  Fig.~\ref{plot1}. 
For most of the parameter space (i.e.~wherever the red lines are below the black lines in Fig.~\ref{plot1}) this constraint implies that the one-to-one correspondence holds.

Another related constraint that must be fulfilled in order that the one-to-one relationship above holds is that the $\phi$ particle, if still stable today and without a vev, is negligibly produced by a possible Higgs portal $H^\dagger H \phi^2$ interaction. Note that if it has a sizable vev, even a very tiny Higgs portal interaction would largely destabilise it. 

At this point, let us revisit the bound on the lightest neutrino mass, Eq. \eqref{numass}. 
If we insist on the presence of a neutrino line, then $m_N > m_\chi + m_\phi > 3.6$ MeV, since Borexino detects $\bar{\nu}_e$ via inverse beta decay, for which the kinematic threshold is $E_{\bar{\nu}} > 1.8$ MeV (see e.g. \cite{Garcia-Cely:2017oco}). 
Combining this with the bound on the DM lifetime from Eq. \eqref{structureformation} and enforcing $\tau_\chi > \tau_U$ gives $m_{\nu_1} \lesssim 3 \times 10^{-6}$ eV, still much lighter than the neutrino mass scale being probed by present experiments.  
Moreover, since Fig. \ref{plot1} typically bounds $\tau_\chi$ to be several orders of magnitude above $\tau_U$, the upper bound on $m_{\nu_1}$ would be correspondingly strengthened. 
Thus, a neutrino line in this model would imply an extremely light neutrino. 
If, on the other hand, we do not insist on the possibility of a neutrino line signal, then a very small neutrino mass could be avoided by taking sufficiently small $m_\chi$, $m_\phi$ and $m_N$.

As already mentioned above, an alternative ``option B" is to consider that the heaviest  particle among $\chi$ and $\phi$ has a lifetime shorter than the age of the universe, i.e.~to consider much larger values of $Y_\chi$. In this case, DM is made of only the lightest species and no neutrino line can be observed. 
A large $Y_\chi$ coupling can change the scenario greatly because it can lead to thermalisation of $N$, $\chi$ and $\phi$. Then the thermalised hidden sector (HS) is characterized by a temperature, $T'$, smaller than the visible SM sector temperature, $T$. Thus one could believe that the one-to-one connection between DM and neutrino mass is lost. This would be the case if later on DM undergoes a non-relativistic, secluded freeze-out in the hidden sector, see  \cite{Feng:2009mn,Chu:2011be}, because in this case the relic density would depend on the annihilation cross section and thus on $Y_\chi$. However, since DM is lighter than the two other particles in the hidden sector, the neutrino portal annihilation processes (for instance $\phi \phi \leftrightarrow \chi\chi$, $N N \leftrightarrow \chi\chi$ or $N N \leftrightarrow \phi\phi$) will in general not decouple when DM is non-relativistic but rather when DM is still relativistic. 
In this case, the relic density doesn't depend on the annihilation cross section but only on $T'/T$ (along the $T'/T$ relativistic floor scenario, see details in \cite{Hambye:2020lvy}). Thus, since $T'/T$ is set by the $SM\rightarrow N$ freeze-in induced by the $Y_\nu$ coupling, here one also finds a one-to-one relation between seesaw parameters and DM relic density. 
	
The value of $T'/T$ can be estimated by considering that at the peak of $N$ freeze-in production, when $T\simeq m_Z$, each $N$ has an energy $\simeq m_Z$, so that the HS energy density is
\begin{equation}
	\rho_{HS}|_{T\simeq m_Z} \simeq n_{N}|_{T\simeq m_Z}\, m_Z=(\pi^2/30)g^\star_{HS}T'^4 \, ,
	\label{rhoscenarioB}
\end{equation}
with $n_N$ given by Eq.~(\ref{N1yield}), and $g^\star_{HS} =9/2$ the number of HS degrees of freedom (from $N$, $\chi$ and $\phi$). Plugging $T'/T$ in the relativistic floor equation for the relic density (Eq.~2 of \cite{Hambye:2020lvy}, see also Eq.~9 of \cite{Hambye:2019tjt}) gives
\begin{eqnarray}
	\Omega_{DM}h^2 &\simeq&c \cdot 10^{18} \,\Big( \sum_i |Y_{\nu i}|^2 \Big)^{3/4} \nonumber \\
	&&  \cdot \,g_{DM}\,\Big( \frac{1\,{\rm GeV}}{m_N}\Big)^{3/2} \Big(\frac{m_{DM}}{100\hbox{ MeV}}\Big)  \, ,
\end{eqnarray}
where $c$ is equal to $2.5$ when determined from Eq.~\eqref{rhoscenarioB} and $9.5$ when properly determined from the energy transfer Boltzmann equation setting $\rho_{HS}$, see Eq.~(49) of \cite{Chu:2011be}. 
This requires slightly smaller values of $Y_\nu$ couplings than in Eq.~(\ref{Omegafreezein}), because the HS thermalisation process increases the number of DM particles. This also implies $m_{\nu_1} \lesssim 8.8 \times 10^{-14}\,\hbox{eV} \,(m_N/1\,\hbox{GeV})\, (100\, \hbox{MeV}/m_{DM})^{4/3}\,(1/g_{DM}^{4/3})$. 
Further details, including the fact that relativistic decoupling in this case requires $m_N\lesssim m_Z/100$ and $m_{DM}\lesssim m_N/10$, are left for another publication. 


\section{Symmetries}
	
The setup assumed above is compatible with various types of symmetries. All symmetry patterns we will consider below assume new symmetries under which both $\chi$ and $\phi$ have non trivial charges while all SM particles, as well as the right-handed neutrinos, are singlets. Thus, the general structure is one of two well-defined sectors, the SM visible sector and a dark sector, each containing particles that are singlets of the symmetries of the other. Both sectors communicate through the right-handed neutrinos, which are natural particles to couple to both sectors, being singlets of each. The dark sector may of course contain more than just $\chi$ and $\phi$. The assumption that DM is created from SM particles through freeze-in can be easily justified on the basis of such a general pattern. One need just assume that the inflaton ``belongs'' to the visible sector, so that reheating proceeds into this sector.

\underline{$\mathbb{Z}_2$ discrete symmetry}: the simplest possibility to justify the existence of a neutrino portal interaction, without  other interactions that could induce DM decays that are too fast, is to assume a discrete $\mathbb{Z}_2$ symmetry under which $\chi$ and $\phi$ are odd, with all other particles being even. In this case, if $\phi$ doesn't develop any vev (so that a possible Higgs portal interaction doesn't destabilise it), the heavier particle of $\chi$ and $\phi$ slowly decays to the lighter one through seesaw interactions, as discussed above in both scenarios A and B.
	
\underline{Global symmetry}: assuming a global $U(1)'$ symmetry under which $\chi$ and $\phi$  have opposite charge doesn't result in any important difference with respect to the discrete symmetry case, as long as $\phi$, which is for this case a complex scalar, doesn't develop any vev. 
	
\underline{Local symmetry}: considering a local $U(1)'$ symmetry under which $\chi$ and $\phi$ have opposite charge potentially induces a number of new phenomena. This will be the case in particular if the scalar $\phi$ breaks this $U(1)^\prime$ symmetry by acquiring a vev, so that the associated gauge boson is massive (as it generally must be). In this case, new extra decay channels for both $\chi$ and $\phi$ arise, which could in particular easily destabilise the $\phi$ state
(for instance by a Higgs portal interaction $H^\dagger H \phi^\dagger\phi$, even if it is extremely tiny).
The $\chi$ DM particle can also be destabilised by new decay channels, in particular from the fact that it mixes with the SM neutrinos proportionally to both the $Y_\nu$ and $Y_\chi$ couplings and to both vevs, $v$ and $v_\phi$, with $\sin \theta_{\nu-\chi} \sim Y_\nu Y_\chi v v_\phi/(m_N m_\chi)$. 
The existence of a dark photon, $\gamma'$, implies a possible $\chi\rightarrow \nu \gamma'$ decay with width proportional to $\sin^2 \theta_{\nu-\chi}$ and the $U(1)'$ fine structure constant, $\alpha'$. It can make the $\chi$ too unstable unless the couplings are small enough or the dark photon is heavy enough.
A $\chi\rightarrow \nu e^+ e^-$ decay is also possible if there is kinetic mixing between the new $U(1)'$ and the SM hypercharge $U(1)_Y$ group. Thus, the seesaw/DM relic density correspondence is viable but requires that quite a number of interactions are tiny.
	
\underline{No symmetries}: If ``just so'' there are no symmetries beyond the SM ones, a number of \textit{a priori} allowed couplings must necessarily be extremely tiny (so as not to destabilise the $\chi$ or $\phi$ DM component(s)), such as $\chi LH$ or $\phi H^\dagger H$ interactions. While possible, this appears more \textit{ad hoc} than with a symmetry.

In summary, seesaw-induced $W$, $Z$ and $h$ decays could be at the origin of the DM relic density, even though DM is not a seesaw sterile neutrino.
Given the current neutrino mass and mixing constraints, the usual type-I seesaw model turns out to have sufficient flexibility to allow freeze-in  production of DM from these decays in a way which is determined only by the seesaw parameters and the mass(es) of the DM particle(s). Two scenarios have been proposed above. As always for freeze-in, these scenarios are not easily testable because they are based upon the existence of tiny interactions, here the seesaw Yukawa couplings of at least one of the right-handed neutrinos, and possibly also the neutrino portal interactions. However, for a whole range of DM masses, scenario A predicts a neutrino-line within reach of existing or near-future neutrino telescopes. Moreover, both scenarios A and B are falsifiable as they predict a small mass for the lightest neutrino.

\section*{Acknowledgments}
This work is supported by the ``Probing dark matter with neutrinos" ULB-ARC convention, by the F.R.S./FNRS under the Excellence of Science (EoS) project No. 30820817 - be.h ``The H boson gateway to physics beyond the Standard Model'', and by the IISN convention 4.4503.15. 
R.C. thanks the UNSW School of Physics, where he is a Visiting Fellow, for their hospitality during this project.

\bibliographystyle{apsrev4-1}
\bibliography{ref.bib}
	
\end{document}